\documentclass[journal]{vgtc}                
\ifpdf
  \pdfoutput=1\relax                   
  \usepackage{graphicx}                
  \DeclareGraphicsExtensions{.pdf,.png,.jpg,.jpeg} 
\else
  \usepackage{graphicx}                
  \DeclareGraphicsExtensions{.eps}     
\fi%

\graphicspath{{figures/}{pictures/}{images/}{./}} 

\usepackage{microtype}                 
\PassOptionsToPackage{warn}{textcomp}  
\usepackage{textcomp}                  
\usepackage{mathptmx}                  
\usepackage{times}                     
\usepackage{cite}                      
\usepackage{tabu}                      
\usepackage{booktabs}                  

\usepackage{multirow}

\onlineid{1224}

\vgtccategory{Research}
\vgtcpapertype{Empirical study}

\title{Untidy Data: The Unreasonable Effectiveness of Tables}

\author{ Lyn Bartram, Michael Correll, and Melanie Tory}
\authorfooter{
\item
 Lyn Bartram is with Simon Fraser University. E-mail: lyn@sfu.ca.
 \item
 Michael Correll is with Tableau. E-mail: mcorrell@tableau.com.
 \item
 Melanie Tory is with the Roux Institute. E-mail: m.tory@northeastern.edu.
}

\shortauthortitle{Bartram \MakeLowercase{\textit{et al.}}: The Unreasonable Effectiveness of Tables}

\abstract{
Working with data in table form is usually considered a preparatory and tedious step in the sensemaking pipeline; a way of getting the data ready for more sophisticated visualization and analytical tools. 
But for many people, spreadsheets --- the quintessential table tool --- remain a critical part of their information ecosystem, allowing them to interact with their data in ways that are hidden or abstracted in more complex tools.
This is particularly true for \textit{data workers}~\cite{liu_making_2018}, people who work with data as part of their job but do not identify as professional analysts or data scientists.
We report on a qualitative study of how these workers interact with and reason about their data.
Our findings show that data tables serve a broader purpose beyond data cleanup at the initial stage of a linear analytic flow: users want to see and ``get their hands on'' the underlying data throughout the analytics process, reshaping and augmenting it to support sensemaking.
They reorganize, mark up, layer on levels of detail, and spawn alternatives within the context of the base data.  These direct interactions and human-readable table representations form a rich and cognitively important part of building understanding of what the data mean and what they can do with it.  We argue that interactive tables are an important visualization idiom in their own right; that the direct data interaction they afford offers a fertile design space for visual analytics; and that sense making can be enriched by more flexible human-data interaction than is currently supported in visual analytics tools.

} 


\CCScatlist{ 
 \CCScat{K.6.1}{Management of Computing and Information Systems}%
{Project and People Management}{Life Cycle};
 \CCScat{K.7.m}{The Computing Profession}{Miscellaneous}{Ethics}
}

\teaser{
   \centering
    \centering
    \includegraphics[width=0.95\textwidth]{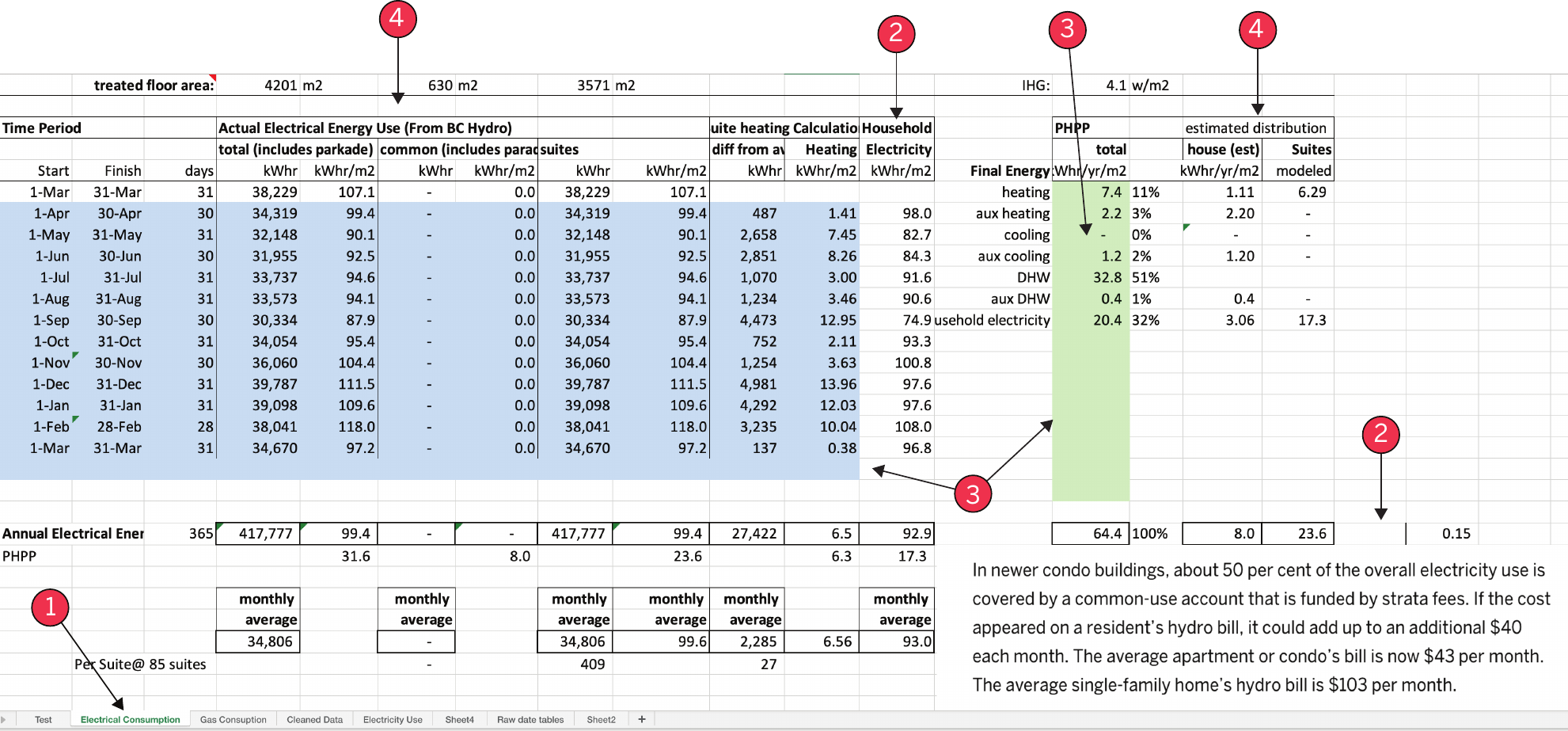}
    \caption{An example spreadsheet (shared with permission of Cornerstone Architects) showing various ``rich'' table features that our participants employed, including (1) A \textbf{Master Table} of \textit{base data} that is often left untouched, with manipulations happening in a \textbf{copy} or other area separate from the base data; (2) \textbf{Marginalia} such as comments or derived rows or columns in the periphery of the base table, often taking the form of freeform natural language comments; (3) \textbf{Annotations} such as highlighting or characters with specific meaning (e.g., a dash denotes missing values) to flag particular cells as anomalous or requiring action; and (4) \textbf{Multi-cell features} such as labels or even data that span multiple rows or columns of the sheet.}  
\label{fig:wound}
}

\usepackage{fancyhdr}
\usepackage{lastpage}
\usepackage[toc,page]{appendix}

\vgtcinsertpkg

\usepackage{subcaption}

\begin{document}


\firstsection{Introduction}
\maketitle

Despite the widespread proliferation of visualization tools, data tables remain ubiquitous. Spreadsheets, the canonical table tools, pervade the data ecosystems of organizations, even when purpose-built business intelligence (BI) tools are widely available. Why? Are these organizations, and their people, simply unsophisticated in their data strategy? Our investigations into data practices (both in this and previous studies \cite{conversationalists}) says emphatically: no! In fact, the table form is an interesting and valuable visual data representation in its own right. Just as data can be ``unreasonably effective''~\cite{halevy2009unreasonable} at solving problems, we observed that tables, oft-maligned by the visualization community in favor of more strictly visual representations, often play an out-sized role in the sensemaking process.
Tables enable users to ``get their hands on the data:'' see it, structure it, organize it, add new meaning, and mark it up in ways that few other visualization tools do. As such, table representations provide a rich and direct means of interaction and an important cognitive structure for thinking with data. Yet these rich interaction possibilities are stripped away when data tables are ``tidied,''  ``cleaned,'' or otherwise ``prepped'' for ingestion into analytics tools. 


We maintain that tables and spreadsheets remain useful across a variety of skillsets and user profiles. We focus on the everyday work that people in business settings, even people who would not self-describe as professional analysts, get done with their data. Following Liu et al.~\cite{liu_understanding_2020}, we term these people \textbf{data workers}. Data workers, in contrast with aspirational ``unicorn''~\cite{bavskarada2017unicorn,davenport2020beyond} data scientists, are found in all sectors, and have diverse levels of data expertise and experience coupled with deep domain knowledge. The work of data workers often expands past boundaries of traditional sensemaking or analysis structures, and encompasses an expansive set of tasks and skills~\cite{crisan2020passing,conversationalists}. Going beyond just analysis, data work is part of \textbf{human-data interaction} (HDI), a broad concept that includes individual and organizational sensemaking activities such as data monitoring, communicating, and munging, and expands beyond visualization to data notifications, spreadsheets, data wrangling tools, and any other means of sensemaking with data. 





The work practices and needs of data workers are poorly understood and ill-supported by current technology, especially with respect to forms like the spreadsheet and the table that may be seen as less prestigious targets for design or research than other visual analytics approaches. Both commercial and academic efforts to support data workers can fall prey to a temptation to fully automate the work (denying data workers the ability to employ their expertise or exercise their agency over the sensemaking process~\cite{heer2019agency}) or focus on narrow groups of users with particular data science skillsets (with the tacit assumption that the mass of data workers will adapt, catch up, or be left behind). Without a deeper understanding of data workers' practices and needs, we risk focusing our efforts on approaches that do not address the core challenges faced in everyday data work. An even greater risk is that we blame the workers themselves, characterizing them as unskilled or employing ineffective work patterns, implying that data workers need to be ``fixed'' through training so they can then work in the same ways and with the same tools as data professionals. Instead, we need to support data workers where they are and empower them to be successful with data.

In this paper, we present the results of a qualitative study of data workers and their relationship \textit{with} and actions \textit{upon} their own data. While our interviews were conceived of as being tool and format agnostic, nearly all of our interviews ultimately centered on the spreadsheet as a common platform of our participants' data work.  We wanted to explore this ubiquity, figure out what makes tables tick, and identify opportunities (or gaps) in existing tooling for visual analytics that all too often leave the table behind. Thus, our original research question of ``how do people who do not describe themselves as professional analysts manipulate, clean, or otherwise munge their data?'' became, over the course of our open-ended exploration, more tightly focused: 
\begin{enumerate}
    \item How do tables, as visual and analytical objects, contribute to sensemaking and understanding?
    \vspace{-2mm}
    \item Which modes, methodologies, and affordances do tables support that other analytical tools do not?
\end{enumerate}




Overall, we found:
\begin{enumerate}
    \item Tables as artifacts have an internal \textbf{richness of meaning} and \textbf{special structural affordances} that are often lost when they are imported or converted for use in other analytical tools: not just annotations and other forms of metadata, but spatial organization, analytical transparency, and modes of interaction that do not survive distillation into the clean and tidy forms 
    demanded by advanced analytics or visualization tools.
        \vspace{-2mm}
    \item \textit{Ad hoc} \textbf{human-data interactions} (like ``eyeballing,'' ``manipulating,'' ``marking up,'' or ``verifying''), despite being both ubiquitous and important in our participants' day-to-day work, are not given their due importance in formal models of data analytics and are not always supported by analytics tools.
        \vspace{-2mm}
    \item Data workers place a high value on building \textbf{trust and understanding} both \textit{from} and \textit{via} their data manipulations. They are willing to perform actions they self-describe as monotonous and repetitive, and face guilt over not using tools they consider more sophisticated or powerful, to maintain immediate access, control, and understanding over their actions and sense making process.
\end{enumerate}

Despite rhetorical/marketing/research shifts away from data manipulation at the spreadsheet level and towards automated- or augmented- analytics tools that generate visualizations, dashboards, and even ``automated insights''~\cite{law2020characterizing} without intervening table generation or inspection steps, the table is an indispensable and inescapable component of how many people do their everyday data work. Tables, and the ways that people work with them, are not juvenilia generated by users who are not sufficiently savvy or sophisticated to use other tools--- a temporary embarrassment that will be dispensed with through a killer app or a comprehensive enough data science education program. Rather, we argue that there are real cognitive and analytical benefits to the table as form; experiences that are not easily duplicated by other ways of working with data: that the table functions as an important (and perhaps inescapable) component of the sensemaking process.
 
We conclude with a call to bring the strengths of interactive tables into visual analytics tools, enabling rich and direct means of getting one's hands on the data. We also encourage researchers to further study the practices of data workers at all stages of experience and data fluency, especially around building trust \textit{in} and control \textit{over} their tools and processes. Existing work that focuses on the idiosyncratic population of professionals may fail to reflect the crucial (but oftentimes tedious, messy, or error prone) data work that people do each and every day. We should think about the best opportunities to intervene as designers in not just providing complex tools to support sensemaking, but also providing the scaffolding that would allow people to build trust, fluency, and confidence in their work.



\section{Related Work}


We first discuss tables in two areas of visualization research: tables used for data wrangling and tables as a visualization technique. We then discuss the work most similar to ours, studies of data work.

\subsection{Table Techniques in Visualization Research}

Tables feature prominently in \textbf{tools for data cleaning, wrangling, and preparation}, a process that accounts for over 80\% of time and cost in analytics \cite{kandel_enterprise_2012}. Tabular data sets present myriad wrangling challenges~\cite{broman_data_2018,rattenbury_principles_2017,murrell_3_2012}. Wickham describes principles for making data tables computationally easy to read, model and visualize~\cite{wickham_tidy_2014}: in a ``tidy'' structure all related columns become levels of an overall dimension and each row is a unique combination of  observations for each level. This results in a ``tall'' format with potentially many rows for each item. Dealing with dirty or ill-defined data introduces additional challenges of cleaning (making data types consistent, ensuring appropriate types), validation (checking for bad data) and removing or replacing anomalous values \cite{azeroual_data_2020, kandel_research_2011}. This may require decisions about densification or imputation \cite{kasica_table_2020, jonathan_drummey_zen_2019} or about what to ignore~\cite{kandel_research_2011}.

Targeted solutions to data cleaning and wrangling 
include programmatic libraries for experts~\cite{mckinney2011pandas,wickham2019welcome} 
but we focus on methods accessible to non-programmers. 
\textit{Automatic} approaches focus on cleaning up and disambiguating data types (see Contreras-Ochando et al.~\cite{contreras-ochando_automated_2020} for a review), interpreting and transforming table structures~\cite{van_assem_converting_2010, chen_automatic_2013}, and using structure to suggest annotations~\cite{de_vos_combining_2017}.
\textit{Semi-automated approaches}
~\cite{kandel2012profiler,kandel_research_2011,raman2001potter} focus on building-block procedures and/or machine-learning assistance related to types and imputation, learning from user input, or context. Commercial examples include Alteryx\footnote{www.alteryx.com}, Trifacta Wrangler\footnote{www.trifacta.com/}, and Tableau Prep\footnote{www.tableau.com/products/prep}. 
\textit{Manual (Instruction)} techniques tackle the problem by changing the work practices of users.  This includes establishing standards on how to reconfigure ``bad data'', including combining datasets into one sheet, structuring tidy tables, cleaning data, and removing common domain habits like markup, formatting and comments~\cite{quartz_bad_data,wickham_tidy_2014,broman_data_2018}.     
 
Bertin~\cite{bertin2011graphics} pioneered interest in the \textbf{table as a visual idiom}, defining the interactive table via the affordances of a fully reorderable matrix. From Bertin's method, Perin et al.~\cite{perin_revisiting_2014} propose critical requirements for tabular visualizations including visual cell encoding, row and column reordering to exploit spatial association and to group meaningful chunks, and annotation of results (e.g., naming groups). 


There is a long history in visualization of adapting table metaphors and affordances for visualizations; e.g., as ``visualization spreadsheets''~\cite{chi1997spreadsheet,chi_principles_1998} or ``spreadsheet-like interfaces''~\cite{jankun2001visualization} where each cell contains a different visualization design, data facet, or both. Tools such as TableLens~\cite{rao1994table}, Bertifier~\cite{perin_revisiting_2014}, and Taggle~\cite{furmanova_taggle_2020} all use the table form as the substrate for embedding visualizations in cells, managing heterogeneous data, and supporting familiar operations such as sorting and rearrangement of rows and columns. Many visualization designers borrow affordances or structures from spreadsheets. For instance, sortable and scannable grids of data appear in recent systems like UpSet~\cite{lex2014upset} and Serendip~\cite{alexander2014serendip}. Gneiss expands the flat table spreadsheet model by embedding hierarchy in cells~\cite{chang_using_2016}. Likewise, the ability to juxtapose data elements, cells, or visualizations with natural language commentary or metadata has been described as an important and powerful tool for understanding in visualization design paradigms like LitVis~\cite{wood_literate_2019}.


Despite the well-established history of tables in visualization, and the utility of spreadsheet-like operations, we argue that tables \textit{per se} are given short shrift in contemporary visualization research--- considered transitory artifacts to wrangle data before analysis or visual representations that require a consistent and tidy data format. For instance, the heatmap borrows the grid structure, but assumes that each cell contains homogeneous data.  
Widely used commercial tools like Tableau\textsuperscript{TM} and PowerBI\textsuperscript{TM} offer table views with limited interaction, intended for data preview or as advanced versions of the well-known pivot table. 


\subsection{Ecological Studies of Data Work}
Numerous studies have explored the practices of people who work with data, most of them focusing on professional analysts and data scientists in enterprise environments \cite{kandel_enterprise_2012,kandogan_data_2014,alspaugh_futzing_2019,liu_workflow-based_2014,muller_how_2019,zhang_how_2020,liu_making_2018,russell_simple_2016}. While most of these studies focused on analysis activities, some characterized additional aspects of data work including data capture, discovery, wrangling, modeling, and reporting, among others~\cite{kandel_enterprise_2012, alspaugh_futzing_2019, muller_how_2019, zhang_how_2020}.

Fewer studies have extended this effort beyond the analytics settings proposed in earlier models (e.g., the Pirolli-Card sensemaking loop~\cite{pirolli_sensemaking_2005}) and to a wider population of people who work with data in more informal, unstructured, or ad hoc contexts~\cite{conversationalists}. 
Bigelow et al.~\cite{bigelow2014reflections} found that designers working with data frequently switched tools and followed a non-linear sequence of tasks. Convertino et al.~\cite{convertino_self-service_2017} reported that ``casual data analysts'' tend to follow a workflow of ``clean and report,'' spending most of their time preparing and organizing data in spreadsheets and struggling to take full advantage of their data. Boukhelifa et al.~\cite{boukhelifa_how_2017} documented three strategies data workers employ to manage uncertainty in their data: ignore, understand, and minimize.
Most similar to our work because of the focus on spreadsheets, Dourish~\cite{dourish_spreadsheets_2017} examined collaborative ``spreadsheet events'' through a sociological lens and explored how the grid structure of the spreadsheet supports particular patterns of interaction. Likewise, Sarkar and Gordon~\cite{sarkar_how_2019} found that people acquire spreadsheet knowledge informally, opportunistically and socially. In particular, like generates like: because people learn from others’ examples, visible features, methods, structures and design practices percolate better than hidden functions.
Many of these studies identify spreadsheets as an important tool for data workers.
Our research extends existing knowledge by focusing on the particulars of how data workers used the form and affordances of tables in their sensemaking practices.








\section{Spreadsheets and Rich Tables}
\label{sec:richtables}
Once tables emerged as the focus of our analysis, we conducted a literature survey about table use. Here we synthesize those findings, forming a frame for analysis of our study. Our observations (section~\ref{findings}) confirm and elaborate many of these themes.

The quintessential workplace data tool is the spreadsheet~\cite{davies_theres_2013,erete_storytelling_2016,ragsdale_spreadsheet_2014}, designed as a multipurpose tool for data entry, structuring, storage, and analysis~\cite{broman_data_2018}. A 2019 survey reported that 88\% of companies use more than 100 spreadsheets in critical business processes~\cite{forrester_study_2019}. The simplest table consists of a grid where rows represent unique items, columns represent attributes, and cells contain values. 
The spreadsheet provides a range of functions to support data modeling, calculation, organization and visualization. However, the majority of spreadsheets are used for storing and interacting with data~\cite{chambers_struggling_2010,handel_what_2016,hermans_enrons_2015,kandogan_data_2014}, often with only the simplest calculation functions~\cite{hermans_enrons_2015}. For many data workers, the spreadsheet's value is as a sophisticated table editor and archive, supporting functions of data integration, manipulation, shaping, editing, annotation and reporting, with minimal calculation~\cite{dourish_spreadsheets_2017,hermans_enrons_2015, ragsdale_spreadsheet_2014}.  In this context, the spreadsheet is the computational tool that instantiates two critical constructs in our work: \textit{untidy data} and \textit{rich tables}.
 
 \subsection{Characteristics of Rich Tables}
 Spreadsheet data organization is table-based but loosely constrained; as a result, real world spreadsheets can be computationally difficult to use verbatim in statistical and analytical tools~\cite{broman_data_2018,murrell_3_2012,rattenbury_principles_2017,wickham_tidy_2014}. Analytical tools might rely on (or attempt to parse) data into a single expected format (e.g., ``tidy data''~\cite{wickham_tidy_2014}), whereas spreadsheet structure is constrained only by the grid, which affords all sorts of spatial organizations or schema. As per Dourish~\cite{dourish_spreadsheets_2017}, ``the grid is fundamental to the materiality of the spreadsheet form; much of what we manipulate when we manipulate spreadsheets is actually the grid—navigating, extending, inserting, deleting, and merging grid columns and rows.'' 

When tabular artifacts are inextricably enmeshed with human sensemaking activity, we refer to them as \textbf{rich tables}. We characterize rich tables functionally, through the presence of one or more of the following features (see Fig.~\ref{fig:wound} for an example):

\textbf{Layout}:
The data layout is structured for human readability rather than machine parsing~\cite{chang_using_2016,murrell_3_2012,wickham_tidy_2014}. E.g., the data might be structured in a ``wide'' (rather than ``tall'') format, where rows contain unique items  
and the attributes of each item are stored in columns. Levels of a dimension are expressed in different columns. This results in a compact, \textit{item-centric} view that is easy to scan~\cite{rao1994table}. Hierarchies occur in both row and column groups where an extra row or column contains a second dimension name~\cite{chen_automatic_2013}.

\textbf{Heterogeneity}:
Cell types are rarely consistent~\cite{lex_visbricks_2011}. Cells within a column may vary in semantics (data meanings), data types, value ranges, and data formats. Columns may have inconsistent distributions, behaviours, clusters and outliers.

\textbf{Missing or sparse data}:
Tables frequently have empty space at a row, column or cell scope, evidence of data quality issues (missing or sparse data) or simply values that remain to be filled in~\cite{drummey_creating_2018, dourish_spreadsheets_2017}. 

\textbf{Multiple datasets}:
Data work frequently involves collecting and integrating datasets~\cite{crisan_uncovering_2019,kasica_table_2020,murrell_3_2012} with different schemas. In spreadsheets these are often kept as separate worksheets~\cite{furmanova_taggle_2020} or even files where the relations and dependencies are incompletely specified or even implicit.

\textbf{Multiple data grains}:
Worksheets have multiple levels of detail ``sprinkled about''~\cite{broman_data_2018}, as different tables or within the same worksheet. Moreover, the data can have different grains \cite{murrell_3_2012,jonathan_drummey_zen_2019}: e.g., sums across columns or rows many appear in the same table as raw values.

\textbf{Annotations}:
Spreadsheet tables often contain visual markup, formatting, and spatial organization (aligned columns or rows) that mean something to the user but are lost in machine translation.~\cite{broman_data_2018, wagner2006enhance}. 

\subsection{Using Rich Tables}

If rich tables are ``messy'' \cite{wickham_tidy_2014}, ``dirty''~\cite{azeroual_data_2020,kandel_research_2011} or just plain ``bad''~\cite{mccallum_bad_2012}, why do these practices persist? In reality, rich tables afford many valuable interactions:

\textbf{Multiscale readability:}
Tables support efficient comparison and are good for looking up values~\cite{few_show_2004}.  
One study  \cite{prodromou_students_2015} showed that students effectively used different access patterns (whole table scan, horizontal and vertical reading, and targeted comparison) to develop a deep  understanding of a multidimensional dataset.

\textbf{Spatial semantics:}
Spatial organization is an important visual resource~\cite{martin_reading_2005, pellegrin_how_2000}. 
Spreadsheet users frequently organize their data in semantically related substructures.
This can include ``blocks'' of contiguous cells or specified ``zones''~\cite{amozurrutia_excel_2011}. These subsets may differ in granularity:  it is common to see summary tables (e.g. pivot tables) near the ``source'' data ~\cite{cao_extracting_2017} (see Fig 1) to support synthesis. The layout serves as ``secondary notation'' that conveys meaning to the user~\cite{hendry_creating_1994}. 


\textbf{Direct editing:}
Spreadsheet tables allow data editing directly on the cell, or by adding new rows and columns opportunistically without having to understand formal data methods like joins~\cite{ragsdale_spreadsheet_2014,sarkar_how_2019}. Rich tables allow attention and interaction at varied levels of organizational scale, from scanning the entire table to cell-based focus. Working at the cell level comprises a large part of spreadsheet activity~\cite{dourish_spreadsheets_2017}, in contrast to working on the data in the aggregate: spreadsheet users select cells, do some operation (like a lookup from another cell), and then see their results. Through cell references, transformations and calculations are explicit to users: they can see and touch both the original data, the operation, and the result. Drummey~\cite{drummey_thoughts_2019} calls this ``cell-based thinking'' and claims that transition to analytic tools like Tableau is difficult due to the mismatch between cell-based thinking and Tableau's aggregation-centric model. 

\textbf{Annotation:} Beyond editing data, people also markup and annotate the data as part of building essential context~\cite{kohlhase_context_2015,wolstencroft2011rightfield}, using layout, formatting, color, and comments at various organizational levels~\cite{sarkar_how_2019,van_assem_converting_2010,kendrick_how_2016,de_vos_combining_2017,rao1994table}. Sarkar et al.~\cite{sarkar_how_2019} refer to ``craft practices and design patterns'' that become formalized in templates that reflect business practices, are shared~\cite{dourish_spreadsheets_2017}, and scaffold opportunistic learning~\cite{sarkar_how_2019}.

\textbf{Flexible interactions:}
Direct manipulation is at the heart of the rich table. Flexible rearrangement operations to associate related items or hide temporarily uninteresting ones support fluid thinking~\cite{rao1994table}. Users create new structure using familiar copy-paste, drag and drop operations that operate via direct selection without requiring knowledge of formal data functions. This materiality  -- the digital experience of working ``hands-on'' -- is an essential process in building knowledge about and with data \cite{tanweer_impediment_2016,dourish_spreadsheets_2017,huron2014constructing}, and is grounded in Piaget's foundational principle that we construct knowledge through action and experience \cite{wadsworth_piagets_1996}.

\section{Methods}

The goal of our study was to understand how data workers make sense of their data; in particular, the practices and interactions they employ when engaging in direct data manipulation. 
Prior to our study, two of the authors interviewed data training experts (E1, E2) to understand the data interaction practices and challenges of their diverse clients. These interviews focused our study design towards an analysis of data interactions and how representations and tools supported data workers' practices. 
User-system interactions are influenced by many factors that are not easily measurable or quantifiable \cite{blandford_interacting_2010, mayr_looking_2016,koesten_trials_2017}. This challenge is magnified when the system is the mediating artifact and usability can influence behaviors and outcomes. We therefore used a mixed-methods qualitative approach combining descriptive techniques, semi-structured interviews, and observational walkthroughs to elicit concepts and capture how peoples' data interactions contributed to their formation of understanding. These methods support insights into behavior and interaction processes as well as the knowledge and concepts that underlie them~\cite{mayr_looking_2016,blandford_interacting_2010,novak_origins_2006}. This approach is common in formative user research and cognitive engineering~\cite{sharples_socio-cognitive_2002}.


\subsection{Recruitment}
We broadly sampled data workers from a variety of backgrounds and sectors with diverse data skills. Participants were recruited from several sources: targeted emails (n=3) to previous participants of visualization training sessions offered by Simon Fraser University or who were suggested by expert consultants as representative users. Our recruitment email solicited participants who had experienced challenges working with their data and who manipulated their data directly within or across business intelligence tools as part of their work. Second, we solicited participants (n=9) on User Interviews\footnote{www.userinterviews.com/}, a common recruiting site for user research. We used a screening survey to refine participant selection (see supplemental material). In all cases, we selected participants who identified their data use as one or more of: (1) use data analyses, visualizations or dashboards that others have prepared; (2) use BI tools to ask and answer questions about data; or (3) use BI tools to create visualizations, analyses or reports for use by others.

We explicitly did not define ``business intelligence tools'' in the screener. Applicants who self-identified as advanced analysts, data scientists, IT support, or data administrators were rejected. In addition, the screener asked about the participant's job and sector, the size of their organization, and their perception of the frequency of difficulties. In all cases, we selected applicants who were willing to walk through a current or recent analysis of their data, explaining practices, concepts and challenges as they did so.  Qualified applicants were contacted by email to arrange study sessions.

\subsection{Participants}
{\renewcommand{\arraystretch}{1}
\begin{table*}[tb]
\caption{Interview Study Participants}
\label{tab:interview-participant}
\resizebox{\textwidth}{!}{%
\begin{tabular}{| l | l |l  |l  |l |}
\hline
\textbf{ID}&
  \textbf{Role} &
  \textbf{Sector} &
  \textbf{Org. Size } &
  \textbf{Analytics Tools} \\ 
  \hline
  P1  & Executive Director & nonprofit, culture & 2 - 10 & Google Sheets, Google Analytics, Excel \\ 
  \hline
  P2  &  Practical Law Specialist &  legal services; law practices & 1000+ & proprietary, web scraping, Excel \\ 
  \hline
    P3  & small business owner, online art market&  retail &  2 - 10 &  Shopify, Google Analytics, Excel \\ 
  \hline
    P5  & GIS analyst&  energy; utility services &  201-1000 &  ArcGIS, Excel \\ 
  \hline  
    P6 & Senior financial Analyst&  financial services & 201-1000 &   PeopleSoft, Excel \\  \hline
   P7  & District Health Information Officer/Analyst &  public health & 1000+ &   Tableau Prep, Excel, Tableau \\  \hline
      P8  & Director of Development&  non-profit & 11-50 &   NGP, Excel \\  \hline
      P9  & Senior Management Consultant, small business owner&  financial services & 2-10 &   Google Analytics, Excel \\  \hline
     P10  & Technical Product Manager&  software services & 51-200 &   proprietary, web scraping, SQL Workbench \\  \hline
     P11  & Data Analyst &  non-profit & 51-200 &   Salesforce, Excel \\  \hline
     P13  & Director of Operations &  health services & 51-200 &   proprietary EMR, Google Sheets, Excel \\  \hline
     P14  & Information Analyst &  non-profit management & 51-200 &   Salesforce, UNIT4 Business World, Excel \\  \hline
\end{tabular}%
}
\end{table*}
}
Our participants (\autoref{tab:interview-participant}) consisted of 12 data workers (4 female, 8 male). Ten were based in North America, one in Africa, and one in Europe. Participants represented a variety of economic sectors and organization sizes, from small business owners to government workers. All but two worked with data using more than one tool and from more than one source. None had formal training in data analytics nor data science.  
All non-government participants were compensated with a \$60 USD Amazon gift card.

\subsection{Procedure}
Sessions took place remotely over Zoom. Participants were asked to bring data from a recent analysis. We first explained the motivation, procedure and potential outcomes. We then conducted a semi-structured interview, a sketching exercise, and walkthough, all focusing on how participants work with data in their job.

Interview topics included: how they used data in their work, their organizational practice (are you supporting others with data and/or analysis?  How?), the kinds of data they used; how they made sense of it and/or transformed it for others to use, where they obtained and retained their data, and the tools and methods they used (see supplemental material for the interview script).  

We then asked participants to describe the data and analysis they had brought to the session, beginning with an overview of their goals and major steps that they took or would take for their work. We asked them to sketch their working model of the data as a way of encouraging introspection about their personal data models. We include more details about this visual elicitation in our supplemental materials. 


 
We then moved to the third stage in the study. Each participant brought an example of a current or recent dataset and sensemaking task representative of their work. We asked them to walk through achieving their task goals with these data, showing us which tools they use, which actions they took, and where they faced challenges. In particular, we asked them to talk us through any instances of the following:
 \vspace{-2mm}
\begin{itemize}
    \item How they arranged their data for sensemaking 
steps or places where the data need to be manipulated, re-organized or amended
\vspace{-2mm}
    \item Tasks that were difficult or tedious 
    \vspace{-2mm}
    \item How they understood the journey from the ``raw'' data to the finished result
    \vspace{-2mm}
    \item How they identified and dealt with missing and erroneous data
\end{itemize}

During the walkthrough, the session lead observed the participant's actions, probed for explanation, and confirmed understanding. We concluded by asking participants to discuss how well their current tools and practices helped them explore relationships in the data, and how they matched and/or impeded how they thought about their data. 

We note that in several cases, participants described their full data scope to us but were unable to bring some of these data to the screen due to privacy and confidentiality concerns. This resulted in some sessions where participants told us about more data than we actually saw. However, we were assured that the data approaches shown to us were representative of these other complex data tasks they undertook with respect to tools, practices, and challenges.

\subsection{Data Collection}
Sessions were audio and video-recorded using Zoom. In 2 cases, we also recorded an additional screen when the walkthrough required two participant screens. Audio recordings were fully transcribed. We also solicited sketches and any other supplemental material participants wished to share; participants sent these to the study lead via email. Finally, we took notes during the session. In three sessions, two study researchers were present and took notes; in the remaining 9, only the lead researcher was present as both observer and notetaker.

We emphasized that we were capturing both audio and video recordings plus participant sketches and observer notes. We explained that we did not need examples of data nor the data schema, but that screen recordings could expose some data. We explained that these raw data would be kept confidential, and that any use of participant-supplied content beyond analytical use by the leads would only occur with express written permission from the participant.  The 12 participants reported here consented to these conditions; 3 additional participants withdrew after the interview due to data sensitivity concerns.
 
\subsection{Data Analysis}
A third researcher, who did not participate in study design or data collection, was brought in at the analysis stage to provide an impartial perspective. We analyzed the data qualitatively through thematic analysis~\cite{Braun2006ThematicAnalysis} of transcripts and videos. All three authors contributed to the analysis, taking an inductive and iterative approach to identify themes and examples. Each researcher independently annotated the transcripts and/or videos; emerging themes were agreed upon through discussion.

\section{Findings}
\label{findings}

We structure our findings around three areas of importance: the physical architecture of the tables our participants built and used (the \textbf{Table as Artifact}, \autoref{sec:tables}), the actions our participants took when using these tables (\textbf{Working with Tables}, \autoref{sec:actions}), and how our participants reflected on their data work (\textbf{Obstacles for Data Workers}, \autoref{sec:workers}).

Per our research goals and methodology, we were interested in exploring the experiences of individual data workers. As such, we do not make (and our interview results should not be used to support) strong generalizable claims about how data work \textit{is} or \textit{ought to be} done. Rather, we point to the diversity of the anecdotes and stories as evidence of the richness of the experiences that data workers have with tables. The emerging patterns among their experiences offer opportunities for follow-on ethnographic work and potential starting points to structure how we think about tables and spreadsheets in sensemaking contexts.

\subsection{Table as Artifact}

\label{sec:tables}
Particpants' table structures involved both  data organization and interaction. Only P10 used a table as a temporary view of a data extract; for most participants their table systems were persistent, individually built up over time (P2, P9, P11, P13) or adopted from existing examples (P14, P8, P5, P6). These tables served variably as analytic ``surface,'' archive, resource (with associated feelings of ownership), communication object, and work practice template; tables embodied practices of spatial organization, data shaping, and annotation that could be flexibly assembled and easily manipulated. In this section, we discuss the table genres and structures we observed.

\subsubsection{Base Data and The ``Master Table''}
\label{sec:mastertable}
While data are never ``raw'' (they are always structured or processed in some fashion), we adopt Dourish's useful concept~\cite{dourish_spreadsheets_2017} of \textbf{base data}: the state, structure, and  grain of the data needed by the user.  
Base data were the lowest levels of detail our users worked with, but  not necessarily all the “raw data” to which they had access. The raw data sources varied greatly in both dimensionality and scale (dataset size), from hundreds (P1,P3,P6,P8,P9,P14); thousands (P2,P13,P7); tens of thousands (P5,P11) to hundreds of thousands (P10) of entries. However, base data typically consisted of hundreds of rows, rather than thousands,  often involving multiple tables. Participants constructed their base datasets using one or more of the following:
\vspace{-2mm}
\begin{itemize}
    \item Combining data from different data sources (P1, P3,P7,P13,P14); \vspace{-5mm}
    \item Manually creating the data by entering values and creating categories (P1,P2,P7,P6,P9, P14);
    \vspace{-2mm}
    \item Extracting some or all of a larger corpus or database (P2,P6,P9, P10,P11,P13,P14).
\end{itemize}
The processes of creating and/or populating these base data structures required them to develop knowledge not only of the raw data in their sources but also how to extract the relevant subsets. 
P2, P8 and P10, for example, constructed their base data by filtering  extracts from external systems into a particular organization of columns. P11, on the other hand, explained, \textit{''I prefer to get the [entire] raw master file, and then from there in Excel... I've got my list and I can manipulate the data, however I choose.''} For others, their workflow incorporated multiple tables in separate tabs (worksheets) linked to the base data, typically with lookup functions. These included intermediate summary tables (P1,P6, P9) or simply different organizations of data subsets each targeted at a separate aspect of analysis (P13, P14). 
 
 Having the base-data representation was critical to understanding provenance in their other tables (P1: ''When I am able to manipulate the data or see and track and cross check the data results with the formulas, then it then it's easy for me to feel comfortable'').

Our participants would derive summaries (e.g., pivot tables or filtered subsets), dashboards, or other visual forms meant for reporting out from this base data, which were then separated (in different, non-tabular formats, in different spreadsheet tabs, or visually separated from the base table). Common justifications for this separation were the removal of extraneous details, or details they believed would not be understood by their stakeholders, in service of tailoring the information for specific (usually known) audiences. E.g., 
P6: ``\textit{I want to create something that's easy for them to read}'' and P8: ``\textit{When I share it with my supervisor... she wants a specific subset. I will pull the subset out and send her just a clean subset.}'' While reasoning with high-level summaries or visualizations was rare in the participants' actual data work, these summaries were most likely to appear in reports rather than intermixed with the base data or generated as part of the workflow (P14: ``\textit{what our executive team  wants to see is not that level of granularity. They want to see their roll up information at a portfolio level}'').


The layer of abstraction or distance between the base data and the subsets or summaries that are analyzed or reported out caused some of our participants to express feelings of deep control or ownership over the table per se: e.g. P8: ``\textit{I’m the only one who edits this.}'' This ownership leads to different standards of documentation than in derived reports: P13 ``\textit{I don’t have other people going in to work on the numbers or change formulas, so I don’t necessarily need to show my work... knowing what I did in each of the cells is kind of just} in \textit{me.}'' This ownership meant that different sorts of annotations (see \autoref{sec:marginalia}) often had meanings that were not directly defined in the chart, but instead kept in the head of the data worker.

For some participants (e.g., P11) the base data table comprised their main work space. 
However, a common pattern we observed was a reluctance to directly alter the base data table, even after it had been extracted from some other tool 
(P6: `\textit{`I don’t change the row[-level] data.}''). This reluctance would occasionally result in a ``master table'' or ``master sheet'' containing only the base data, separated spatially or contextually from any derived data or annotations. 
Some participants were hesitant to interact with this master table in any way. This could be for caution (P2: ``\textit{[I am] paranoid that I will do something that will ruin the original sheet,}''), for record-keeping and replicability (P5: ``\textit{I kept all the data intact and included the list so I can go back}''), or because the master table itself is the repository of the data, and so the source of truth (P11: ``\textit{We think of [this workbook] as the data repository in many ways... but if we’re going to be doing any sort of analysis, that would be pulled in elsewhere}).'' 


If the master table is the \textit{sanctum sanctorum} within which no interaction or adornment is permitted, then data work must happen elsewhere, typically on a copy, often in a separate tab within the same workbook. P7 referred to their copy of the row-level data table as ``\textit{the `workout view'...this is the starting point for me.}'' 
P6, with a similar data organization, said ``\textit{From this I can get all the raw information already. But in order to do deep analysis, I have separate tabs.}''
These staging grounds allowed our participants to explore, experiment, and verify their work. For instance, P1 used explicit links to a master table in a formula to build confidence in their work: ``\textit{when I am able to manipulate the data or see and track and cross check the data results with the formulas, then it then it's easy for me to feel comfortable}.''

One challenge was making comparisons, particularly when a master table plus variants was used to explore alternatives. For example, P10 created multiple sets of simulated search engine results, then found it challenging to compare those what-if scenarios. Even during our interview they got confused about which data were which: ``\textit{But by having multiple tabs open, simulating the same search with tweaking some parameters, I got lost. So this is an area of breakdown... not knowing the relationship between the simulated and the actual.}''

\subsubsection{Spatial Organization and the Readable Table}

We turn now to how our participants shaped their working tables.  Fig. \ref{fig:wound}, from a pilot interview, 
illustrates some of the ways our participants built and adapted rich tables.
These patterns illustrate how the table functions as pedagogical structure, visualization, and directly manipulable object for many of our participants.
The tables our participants used were all in ``wide'' format with one row per item. For those who explicitly tried to refactor their data  into ``tall'' format for other tools, such as P1 and P7, this violated their model of how to think about the data. Our participants all had an \textit{item-centric} working model of their data (P1: ``\textit{the individual is the row, and then the columns represent various segments...in my head, it's either that the person or the organization is the row and everything else has to be the column or the reverse}''). This allowed them to read and compare items across multiple columns efficiently (P7: ``\textit{So if we are getting two reports from one facility for a single month, we  know something is wrong, because that's not how they are supposed to report}'').  Wide tables were the default format even for data that was not natively tabular. P10, for example, pulled snapshots of a very large JSON dataset into tabular views to ``\textit{get a sense of my data at a macro scale}'' with respect to detecting key attributes to form subsequent queries.  The wide table was viewed as a natural structure for organizing data views, accommodating multiple levels of detail, and comparing and grouping the effects of filtering and sorting on items. 

Participants used spatial organization to visually associate or discriminate  data both within and across tables, taking advantage of the strong perceptual organizing principle of proximity. These spatial organizations were either fixed (defined at the time of building the table) or dynamic (responding to re-ordering interactions). Some participants (P1, P5, P9, P13) created explicit hierarchies with multi-column headings; these were often used to reorganize the base data into more meaningful segments.  Only one (P9) used a multi-row heading in summary tables.  More common was spatial adjacency for clustering. Common table structures grouped related columns together, ordered where appropriate, and in some cases (P8, P9) representing implicit hierarchy (e.g.,  ``Category, Sub-category'').

Spatial organization also mattered across tables. The  flexible container of the \textit{sheet} afforded different ways to organize combinations of views and summary tables.
P6 grouped pivot tables and filters in tabs specific to different analytical questions of corporate performance -- for both analysis and communication purposes. In contrast, P1 built sheets of summary tables for their own path to the final analysis, with the overall summary at the left and sub-populations to the right. Tables were typically separated by white space and formatting and had multicolumn headers similar to pivot tables.

\subsubsection{Data Granularity}
The wide table format supports easy expression of different data grains; this was important to participants as it helped to ``\textit{cross-reference data and see [results] more clearly}'' (P1).  P7 brought up the challenge of managing different data grains in tidy tables for analytic tools; they used monthly counts (one grain) of Covid cases and yearly populations (a coarser grain) by region to analyse disease spread. In Excel it was easy to sum up the case counts to the regional-level and then divide by population to derive the infection percentage using either a temporary table or direct cell-based calculations, but in Tableau Prep they had to go through a series of painful pivot steps to avoid propagating the wrong data grain. The most common examples of different data grain are summary rows and columns, which were very common in both base and derived tables (observed in the tables of 10 participants). 

A related challenge was structuring and presenting the same data at multiple levels of detail to manage communication. As per P2, ``\textit{
[our VP] just wants information a high level, he's not going to drill down into the data in the same way that say, for example, one of the lawyers who's written the content might do... I find myself creating different versions of a spreadsheet of data, just to share with different people.}'' 
 
\subsubsection{Marginalia and Annotations}
\label{sec:marginalia}
There was an accumulation of marginalia, commentary, and other sorts of annotation, especially for participants who refrained from editing the central master table (\autoref{sec:mastertable}). Our participants variously included comments, color or symbol legends, or just detailed explanations of the data provenance, in the areas around the master table of the base data.

We refer to \textbf{marginalia} as additional information in the periphery of the master table: e.g., derived columns, comments, descriptions, and summaries. Marginalia are designed for human readers, and add context or additional details to otherwise unadorned base data. Natural language descriptions (of the meaning of particular columns or tables, of data provenance, or summarized findings) results in cells in the margins containing text data that is rarely standardized or amenable to actions like sorting or pivoting (P8, ``\textit{[the notes column]'s not really sortable because the notes are kind of freeform. But that’s by design}.'')

A specific type of marginalia are bespoke categories or other derived columns. For instance, P8 manually created categories of contacts: ``\textit{So I've created kind of my own data sorted categories ... they're either VIP, they're not, they're currently a donor, they're a prospect, or we're trying to figure out who they are.}'' 

We refer to \textbf{annotations} as additional information included within the main body of a table, for instance highlighting a cell or changing its value to match a convention. 
Annotations served to flag important parts of the table for follow-up, disambiguating values that would otherwise appear identical, adding codes, noting concerns about data reliability, or providing structure to help navigate the visual complexity.
P14 used dashes in cells that would otherwise be blank to indicate a data quality issue, whereas P2 would highlight an entire row if there was ``\textit{a major red flag.}'' Disambiguating different sorts of missing data was an important task for P5: ``\textit{[a] blank value is telling me that I don’t have a customer of that type in that county.}'' But for cases where the blank value was indicative of a data prep failure (say, a failed join), they used annotations so the final cell was not blank: ``\textit{If something got screwed up, I wanted a placeholder on the data.}'' P8, by contrast, used highlighting for triage: ``\textit{if something is highlighted in yellow then I need to act on it in two weeks.}'' 
Lastly, P13 used color coding to flag key performance indicators over time as poor, satisfactory, or good.

An action that sits at the intersection of marginalia and annotation is in-cell commenting. Curiously, we did not observe any explicit use of the in-cell commenting features afforded by Excel and Google Sheets; participants tended to create comments in separate columns inline with the rest of the data (although often spatially separate from the master table) rather than existing as something to invoke by clicking on a cell. While we did not specifically ask about cell-based comment features, it is possible that the reluctance to use them was based on a loss of ``glanceability'' or the likely loss of these comments when the table is translated to or from other tools.

\subsection{Working with Tables}

\label{sec:actions}

In this section we focus on the actions that our participants performed with their data, especially sensemaking actions (like altering data schema, adding or combining new data, and verifying data integrity) and the extent to which they were supported by tabular forms. Actions that made use of the tabular form and/or the spreadsheet were present across all stages of the sensemaking process, from data preparation and cleaning, to analysis, to reporting and presentation. 
While the table form was more or less strongly represented at these different stages, it was never entirely absent: manual interaction with a table is not so easily dispensed with, despite the focus in data science on repeatable or automated data flows, or best practice guides with statements like ``there is no worse way to screw up data than to let a single human type it in, without validation''~\cite{quartz_bad_data}.

\subsubsection{Moving Between Tools}
\label{sec:moving}

While spreadsheet software was ubiquitous among our participants, only 3 participants \textit{exclusively} used spreadsheets. Participants populated their tables by extracting data from other tools (see \autoref{tab:interview-participant}), such as Salesforce (P2, P3, P8, P11, P14), PeopleSoft (P6), webscrapers (P2, P8, P10), Shopify (P3), or proprietary tools (P5, P8, P13, P14). As such, participants would occasionally \textit{drop in or out} of spreadsheet software for other tasks. Note that for most of our participants, their primary data source was a big data system that included a visual interface to data (e.g. Google Analytics, Shopify), yet they still dumped data out of these tools to work with it in spreadsheets.

One reason to change either tools or table contexts was to \textbf{report out} in a form that would be more useful for external stakeholders. For P9, that meant manually entering in values from a pivot table into PowerPoint: ``\textit{if I were a hands-on CIO I could go in and play, [but] in most circumstances somebody will need to dump out of the analytics tool ...
the client just wants the data presented in PowerPoint,}'' whereas for P6 this was connecting dashboard templates to aggregated data. 
Similar to observations by Dourish~\cite{dourish_spreadsheets_2017} on the ``performative aspect'' of the spreadsheet, often a table \textit{per se} was the medium of external communication (although often simplified or filtered for external audiences). For instance, P5 claimed that ``\textit{what I do with that is dump it into Excel, put it into a pivot table, and distribute it out}.'' 
P13 communicated with their stakeholder (a CEO) with a pivot table, but admitted to looking for ``\textit{a better, easier space for someone who's not a math nerd.}''

In other cases, the move into spreadsheets was temporary, with the table acting as an interlocutor between software tools. For instance, P14 used Excel as a bridge between Salesforce and their other business data management system: exporting from Salesforce, manually joining (see \autoref{sec:joins}) and augmenting the data with information from their other sources, and re-importing back into Salesforce. 
The \textit{ease of use} of working with tables was frequently cited as a reason to export to tables from other systems: P9 was ``\textit{more comfortable}'' in Excel, P2 called Excel ``\textit{pretty much our best friend,}'' and had ``\textit{a lot more confidence speaking on how this data is being pulled [in Excel],}'' whereas P11 admitted ``\textit{I don't really like working with Salesforce in terms of data manipulation... I prefer to get the raw master file in Excel.}''

Participants occasionally used other data cleaning tools, but defaulted to spreadsheets when they encountered barriers. 
P1 reported: 
``\textit{I can do manual cleaning, and generally it's okay, it's just time consuming. And it felt like it was taking me more time to create the formulas to clean it in [Tableau] Prep ...
Google Sheets is my kind of my world.}'' 
We explore this notion of guilt over tooling choices further in \autoref{sec:guilt}.

\subsubsection{Eyeballing \& Verification}
\label{sec:eyeballing}

A common task we observed with tables was ``eyeballing'' the data to confirm that they met expectations. 11 of our participants made explicit statements about verification or other forms of data quality assessment. As per P8, ``I'm \textit{the quality control.}'' Despite the importance of verification, two participants, P3 and P7, were clear-eyed about ``perfect'' data as being an unattainable goal: P7 said ``\textit{Data is never perfect... [if you don’t look at the data, then] the data will surprise you at the moment that you least expect.}'' whereas P3's motto was ``\textit{I don’t want to put garbage in: I want to have the best garbage possible.}''

Eyeballing was often \textit{ad hoc}, rarely connected to formal data cleaning procedures or algorithms (per P3: ``\textit{It's not science per se... just visual}''). An exception was P5, who would ``\textit{run that data check and get an exception report}'' to confirm that, e.g., all 2-inch valves were connected only to 2-inch pipes. Rather than formal procedures, we observed participants encountering ``\textit{a discrepancy}'' (P1,P11),``\textit{alarm bells}'' (P2), ``\textit{major red flags}'' (P2), ``\textit{something out of the ordinary}'' (P8), ``\textit{something [that] drastically changes}'' (P9), or failing a ``\textit{spot check}'' (P5,P10). Participants would then dive into the base data to diagnose errors (per P9: ``\textit{after analyzing the data three or four times, you notice a pattern... if something drastically changes, then you want to double check at the Excel level}''). Eyeballing happened at all stages of data work, not just during data preparation; for instance, P14 encountered an incorrect value in the pivot table intended for presentation to stakeholders \textit{during our interview}, and made a verbal note to update this value once the interview was over.

For some participants, eyeballing occurred in a (master) table of the base data. For instance, for P6 ``\textit{When I validate my data, normally, I always go to the row[-level] data}.'' Directly looking at the base data provided participants with a sense of control over their data plus direct access to fields to be corrected. As per P2: ``\textit{[the spreadsheet] offers me the flexibility to kind of eyeball my data and manipulate it into a manner that is sufficient for me}''
and for P10, ``\textit{I can read [the base data] and understand what this means... and if there's any data missing, or if data is not being parsed out correctly, I can catch that with some spot checking... like a kind of quick and dirty way to poke around.}'' P9's eyeballing could encompass the entire dataset:  ``\textit{if something is out of the ordinary then I double check thousands and thousands of rows.}'' But this was relatively uncommon. P10, for instance, would create \textit{ad hoc} subsets of the base data in order to perform closer inspections, or aggregate the data to a higher level of detail before eyeballing.

As a way of counteracting the issues of scale like those encountered by P9 and P10, other participants would focus only on the data that had been updated or changed. E.g., P6's workflow was ``\textit{I don’t blind upload anything... I keep multiple tables within that Excel sheet of people that need to be added... and I screen that extensively [because it is] hard to go back and figure out where and when exactly new data came in.}'' Perhaps more formally, P5 stated that ``\textit{We do a daily inspection... one analyst every day takes a global look at all the changes for that day and really just does a spot check,}'' and stated that this standard was being augmented by bringing in an exterior data cleaning team: ``\textit{We've kept a dirty house for a while, and now we're getting it all cleaned at once.}'' Interestingly, P5 continued to perform data cleaning operations in advance of the arrival of this professional data cleaning service; rather than redundant, we suggest that this action might be part of deriving a feeling of ownership over the data, in the same way we might clean our homes more if we know strangers will be coming over. 

A related rationale for checking changed data was a lack of trust in the the worker's own actions (but an implicit or explicit trust in the base data). E.g., P14: ``\textit{it’s really heavily manual. So that means that there’s always the concern about data quality from the inputter, which is predominantly me.}'' and P9 ``\textit{I don’t have to doubt the data that Google is providing... what I doubt is the analytics that I do in Excel.}''

\subsubsection{Manipulating Rows and Columns}
While it was uncommon for our participants to insert columns within the base table, it was common for them to append new columns as marginalia (occasionally with blank columns as visual separators between base and derived data). Exceptions were P2 and P11, who created and inserted new calculated columns adjacent to the source column. 

Most participants did not dynamically reorder columns in their tables, largely due to lookup dependencies (P1, P9, P14).  Some whose main workout space was the base table, however, relied on dynamic ordering. P2 frequently rearranged columns to get a single page snapshot of what was most relevant: ``\textit{I need to see the most important information first}''. P11 reordered columns to  collect related data after updates, as  results appeared in a new, ``distant''  column.
However, many participants hid and revealed columns to create visual subsets, supporting different aspects of sensemaking and communication.  For example, P14 used hiding to focus  attention on specific sub-organizations, so they could drill-down or rollup detail columns. Hiding columns was a common strategy to manage very wide tables (P2, P8, P11, P14). 

The most common spatial organization strategies used filtering and sorting to group associated rows (eight participants).  Sorting was also a key strategy in adding data to the table using column alignment (see \autoref{sec:joins}). For those who worked in the base table, filters and sorting were key tools to segment and relate items. 
Participants valued the ability to combine filters via menu options, and to have them dynamically update with the data (P8: ``\textit{when I add a new category [type], it automatically gets added to the list, so I can always see the options}''). P3 and P10, by comparison, relied primarily on sorting to determine the range of values in relevant columns. 

\subsubsection{Re-coding and Joining}
\label{sec:joins}

We saw that participants frequently re-coded their data, either by adding new types in existing columns (P2,P8,P13,P14) or by defining new derived types (P1,P2,P5,P9,P13,P14). An interesting example of re-coding was P1. Working with survey data, P1 needed to report on gender demographics as part of a diversity and inclusion initiative while also (as per accepted practices for inclusive gender solicitation in web forms~\cite{jaroszewski2018genderfluid}) providing free-text responses for gender presentation. They attempted to group gender identities using regular expressions, but were unable to write a regular expression that was specific enough to exclude ``Woman'' but include other text that ended in ``man.'' As such, they manually placed ``Woman'' with ``Womxn'' to make their regular expressions easier to write and their analytical questions about gender easier to answer.

While participants felt free to add to, annotate, re-code, or derive new information in their local data, they rarely adjusted 
data schema in source repositories.  P5, when asked about whether or how they would perform a batch update, claimed that such a task would ``\textit{involve the IT department and [they'd] say `hey, run this script'}.'' An exception was P13, whose workflow involved updating not just data but also the schema to match new (and more convenient) standards. They would locate the deprecated data in Salesforce, export to Excel for manipulation, schema correction and cross-checking, and then perform a batch update back to Salesforce--- a tedious and manual process (``\textit{my job is doing that over and over and over}'').

A related problem (and persistent challenge) was joining together multiple sources of existing data: (P5: ``\textit{---that’s really the heavy lift,}'', P9:``\textit{was such a pain that I decided to do away with it.}''). This was the case even for tools like Tableau Prep (P7) that have visual interfaces for joining tables (albeit ones that require an understanding of database concepts such as inner, outer, left, and right joins). Instead, participants used tables for joining or ``bridging'' disparate data sources in their working tables. Some used formal lookup functions (P1,P11,P14). Others used a shortcut solution to the difficulty of performing (or understanding) data joins afforded by the table structure: the \textbf{copy and paste join} (P7, P9, P14). That is, the user sorts or pivots both sheets based on a common field, then simply copies and pastes fields from one sheet to the other. Both P7 and P9 admit to employing copy and paste joins in their workflows: as per P7, ``\textit{I don’t always get it right. The joins [in Tableau Prep] are quite tricky, [whereas in Excel] it’s really simple and straightforward: you just have to copy one column: you just have to copy and paste, you don’t have to worry about the joining columns and all that}'' (although they did admit some guilt about such manual and error-prone procedures, see \autoref{sec:guilt} and \autoref{sec:moving}).

\subsubsection{Manual Calculations \& Manipulations}
\label{sec:formulae}

Strangely absent in the tables we observed 
was the heavy use of formulae. Formulae we did encounter were relatively straightforward (for instance, running totals of rows or columns, or lookups to link related tables). Occasionally these formulae were for the benefit of readability rather than mathematical utility. E.g., P6 had a running total column, but only so the total would be ``\textit{more visible for people to get the result right away rather than to do the math in their mind}.''

An interesting example of wanting control and ownership over formula evaluation was P13, a self-described ``\textit{calculator junkie,}'' who used a handheld calculator to create projections for their metrics of interest (P7 also used a handheld calculator as a supplement to their table, but for ``eyeballing'' and verification purposes, see \autoref{sec:eyeballing}).
Although they were aware that using the calculation engine in Excel could support transparency and reproducibility, they felt their comfort with their calculator and their sole ownership of the table data justified their exclusion: ``\textit{I don’t have other people going in to work on the numbers or change formulas, so I don’t necessarily need to show my work}.'' Another example of eschewing certain forms of computational automation was P1, who manually built what was functionally a pivot table in Excel, cell by cell, through manual lookup commands. Their stated reason for doing so (rather than relying on built-in pivot table functionality) was ``\textit{[I want to] easily see and track/crosscheck the data results with the formulas. Then it's easy for me to feel comfortable with what’s being reported.}'' That is, they were relying on the formulae highlighting in Excel to show precisely which numbers from the master table fed into the resulting pivot table, for verification.


\subsection{Obstacles for Data Workers}
\label{sec:workers}

Soliciting challenges that participants encountered in their data work revealed a recurring tension between the tedium of manual work practices and barriers to adopting new tools.

\label{sec:tedium}

A consistent complaint was that tasks were often rote, repetitive and boring, and yet still required manual action and attention. P14 said ``\textit{it's just the tediousness of it that makes it difficult, and the manual data input, which always leads to the opportunity for error}'' 
while P11 complained about doing the same complex re-coding task ``\textit{over and over and over}.'' 
P9 even admitted to giving up on more sophisticated data work (like joins, see \autoref{sec:joins}) in the face of workload: ``\textit{I let the people in my organization know to [settle for a lower] level of insight because it just takes too much time... I wish Google Analytics just} did \textit{it.}'' P13 was confident in doing their work with the current scale of their data, but was worried about the future: ``\textit{the point is pain avoidance... as we continue to grow, it’s only going to get worse}.'' 



\label{sec:guilt}

Our participants were cognizant of the extra manual effort (and opportunity for error) of performing their data work exclusively in spreadsheets. As such, they occasionally expressed a desire to use or master other tools, at times resembling \textit{guilt} over not using more ``sophisticated'' tools. P13 stated ``\textit{I haven’t [used visualization tools], but it’s something that I need to.}'' P10 likewise stated: ``\textit{I achieved a proficiency in doing data analysis in spreadsheets ...
sometimes it was easier to perform a specific task in [a spreadsheet], as opposed to maybe a more elegant solution of actually using SQL or a BI solution.}'' Another barrier to adopting tools was a sheer lack of time or background knowledge. P11 stated ``\textit{I feel like this would be a lot easier if I knew SQL... But you know, at a nonprofit, everything's always on fire. So taking time to learn SQL is not generally on my schedule.}''

P7 tried particularly hard to learn new tools. Despite struggling with joins (see \autoref{sec:joins}), P7 felt an obligation to remain in Tableau Prep: ``\textit{I know it’s easier for me to go back to Excel and do one or two things, but I try to persevere ...
I want to see how far I can get with Tableau Prep before I have to give up.}'' Building fluency (for future benefits) was an important goal for P7: ``\textit{If I stick to Excel, then I will have to spend a lot of time doing something that is really really simple.}'' 
P7 employed an iterative process in Tableau Prep in order to build comfort and verify accuracy: ``\textit{you want to have very simple building blocks and move slowly.}'' After each atomic step (such as a join or pivot), they would validate values (occasionally using a handheld calculator) to ensure they were getting ``\textit{the actual figures [they] expect}.'' 

\section{Discussion}
Our exploration into the activities of data workers led us inexorably to an investigation of how people use tables to make sense of their data, organize and structure their workflow, and document their findings. The spreadsheets we saw were not mere matrices of values. Instead, our participants constructed, used, and presented their tables in rich and sophisticated ways. 
We found that many data workers rely on tables as a trusted window with ``direct'' access to their data in a form that gives them both precise control and a sense of confidence and ownership over the data manipulations they want to perform.
Whether as the original data source, the ``home base'' for sensemaking, a ``work space'' for testing and verifying data manipulations, or an intermediate or interlocutory artifact, spreadsheets were ubiquitous elements in how data workers got their work done. The form of the spreadsheet, in turn, shaped the sort of work people were doing.

We focus our discussion on the structure, uses, and unsolved challenges in the genre we call \textbf{rich tables}.
The predominance and persistence of data work involving rich tables has concrete implications both for how we as designers ought to support data workers and how we as researchers should conceptualize the people and processes we study.

Rich tables are living, changing data documents that include annotations, comments, and spatial structures that are important for understanding, reading, and presenting the data. Analytics tools often assume tidiness in the input data (e.g., few could accept something like \autoref{fig:wound} without hiccups). Rather than focusing on ``tidying'' the structure or ``cleaning'' the data to fit a proscribed machine-readable form, designers should \textbf{support a wider variety of human-legible metadata} when modeling, importing, or working with data. For example, permitting the native wide format,  preserving both explicit and implicit associations from spatial structure such as aligned columns, linked tables and summary views, or including user-generated markup and marginalia. An example of affording flexibility in data structures is the Falx system~\cite{wang2021falx}, which attempts to infer ad hoc data transformations in order to generate custom visualizations rather than working around an assumed single and immutable data structure.

The prominence of untouchable master tables suggests value in \textbf{giving users areas for consequence-free and transparent experimentation with data}. Experimentation is especially important for verification tasks related to calculations, joins, and aggregation; many of our participants were willing to perform extremely labor-intensive processes rather than rely on automatic operations about which they lacked trust or confidence. To support scaffolding and ``debugging'', we might need to \textbf{augment visualizations with explicit data flow information} as suggested by Hoffswell et al.~\cite{hoffswell_augmenting_2018} or \textbf{explicitly visualize how calculations and aggregations lead to a final visual result} as per Kim et al.~\cite{kim2019designing,kim2020gemini}. We could also offer no code mechanisms to explicitly support what-if analysis and comparison of alternative scenarios (e.g., P10's multiple search simulation runs).

Our participants continuing to do their work in spreadsheets, despite knowledge of (or guilt over not using) more ``sophisticated'' BI or visual analytics tools points to a need for \textbf{\textit{scaffolding} and \textit{support} for new users}. To integrate other tools into their workflow, data workers need to be able to trust and rely on them in the same way they currently do with spreadsheets, and be able to seamlessly fit them into their workflow in the face of time and resource pressure. Note that this is not just a matter of education; because data is not necessarily the core of their job, we need to develop \textbf{fundamentally different means of interaction} to support data workers. Ultimately these approaches will benefit everyone, data workers and professional analysts alike.



Our most salient finding was the sheer irreplaceability of the table form. There is a temptation in analytics tools to focus quickly on abstractions (e.g., overview visualizations). While abstraction and summarization is powerful, \textbf{if people can't get direct access to the base data in your analytics tool, they'll leave it and go to a tool where they \textit{can}.} The power and immediacy of being able to see, manipulate, and validate data was critical to our participants' data work, something they missed when attempting to use non-tabular tools, and occasionally something they were willing to endure tedium or extensive effort to retain.
Despite being a workhorse of data work, the table has received short shrift in contemporary visualization research. While books and blogs for non-academic audiences have many guidelines for the construction of tables (e.g., Few~\cite{few_show_2004}), definitive research on the design and use of tables as \textit{visualizations}, or embedded within \textit{visual analytics processes} is relatively sparse. We call for \textbf{a renewed focus on the table alongside other visual forms of data communication}.

This neglect has implications even for our empiricism: Joslyn \& Savelli~\cite{joslyn2020visualizing} point out that empirical studies in visualization often lack textual or tabular conditions. When do users really need novel or complex visualization systems? Sometimes, a simple table may suffice, or perhaps even outperform, as in early experiments by Carter~\cite{carter_experiment_1947}.

Lastly, we note the focus in visualization research on ``domain experts'' and ``analysts''. While these groups provide useful new problems and datasets, they are not representative of all (or perhaps even most) data workers. We strongly encourage researchers to \textbf{explore work practices of all types of users, in the wild}. Millions of data workers are using data every single day. They are employing complex and tedious workflows with spreadsheets rather than wholeheartedly adopting the visual analytics tools we work so hard to produce. There is an enormous opportunity for visualization to become the enabling force for data driven decision making in organizations. But to do so, we need to fundamentally change who we, as a research community, conceptualize as the visualization user and the skills they possess or prioritize. Rich interactive tables offer one way to bridge that gap.

\subsection{Limitations \& Future Work}
Our research is exploratory and formative, and consequently limited in scope. In many ways our contribution is not only to report on findings of what users do but also to surface questions to be addressed in future work. Notably, our study focused on self-reported behaviors, so we did not specifically question participants  about their cognitive processing. Extracting and defining  mental models is a difficult task \cite{carley_extracting_1992}. Further, we did not explicitly probe users for how the spatial organization and reorganization of their tables helped them develop a mental model nor how the organization aided their processing and workflow sequences -- but it exposed these practices in common use. We encourage future work to explore what spatial structures, and the acts of manipulating and redefining them, might mean for cognitive processes and mental models. Our future work will also examine how features of rich tables, such as marginalia, may support both individual sensemaking and collaborative workflows \cite{dourish_spreadsheets_2017}. 

\section{Conclusion}
We examined the sensemaking practices of data workers. Describing this process is one and the same as describing the genre of \textbf{rich tables}, the tabular, human readable and human manipulable format that was ubiquitous in the data work we observed. While some (including some of our participants!) disparage the table or spreadsheet as a primitive analytical tool to be inevitably replaced by sophisticated data management or analysis tools, it persists all the same. As a tool for allowing someone to immediately ``see'' and ``get their hands on'' the data, the table has affordances that are unmatched even by complex analytical systems. We should strive to bring those affordances to the visual analytic systems of tomorrow.

\acknowledgments{
This work was partly supported by the Natural Sciences and Engineering Research Council of Canada. We thank Jonathan Drummey, Bethany Lyons, Maureen Stone, and the reviewers for feedback.
}


\bibliographystyle{abbrv-doi}

\bibliography{relatedwork.bib}
\end{document}